\documentstyle[12pt,aaspp4]{article}
\def\icm{${\rm{cm}}^{-1}$}

\def\he3{$^3{\rm He}$}

\def\ROSAT{{\sl ROSAT}}
\def\GINGA{{\sl GINGA}}
\def\IRAS{{\sl IRAS}}
\def\ao{{Appl. Optics}}

\def\etal{{\em et~al.}}
\def\beq{\begin{equation}}
\def\eeq{\end{equation}}
\def\uK{\hbox{$\mu$K}}
\def\arcmin{'}

\slugcomment{Submitted to Ap. J. Letters}
\lefthead{SILVERBERG ET AL.}
\righthead{Millimeter/Submillimeter Search for the 
Sunyaev-Zel'dovich Effect in Coma}
\begin{document}

\title{A Millimeter/Submillimeter Search for the 
Sunyaev-Zel'dovich Effect in the Coma Cluster}

\author{
R.~F.~Silverberg\altaffilmark{1},
E.~S.~Cheng\altaffilmark{1},
D.~A.~Cottingham\altaffilmark{2},
D.~J.~Fixsen\altaffilmark{3},
C.~A.~Inman\altaffilmark{4},
M.~S.~Kowitt\altaffilmark{1},
S.~S.~Meyer\altaffilmark{4},
L.~A.~Page\altaffilmark{5},
J.~L.~Puchalla\altaffilmark{4},
and Y.~Rephaeli\altaffilmark{6,7}}
\altaffiltext{1}{Laboratory for Astronomy and Solar Physics,
NASA/Goddard Space Flight Center, Code 685.0, Greenbelt, MD
20771}
\altaffiltext{2}{Global Science and Technology, Inc., Laboratory for 
Astronomy and Solar Physics, NASA/GSFC Code 685, Greenbelt, MD 20771}
\altaffiltext{3}{Hughes STX Corporation, Laboratory for Astronomy and Solar 
 Physics, NASA/GSFC Code 685, Greenbelt, MD 20771}
\altaffiltext{4}{Enrico Fermi Institute, University of Chicago, 5640 S. Ellis Avenue,
Chicago IL 60637}
\altaffiltext{5}{Princeton University, Physics Dept., Jadwin Hall,
Princeton, NJ 08540}
\altaffiltext{6}{Center for Particle Astrophysics,
University of California, Berkeley}
\altaffiltext{7} {School of Physics \& Astronomy, Tel~Aviv University, 
Tel~Aviv, Israel}

\authoraddr{correspondence regarding this manuscript should be addressed to: \\
            Robert Silverberg \\ 
            NASA/GSFC Code 685 \\
            Greenbelt, MD 20771 \\
            Phone: (301) 286-7468 \\
            FAX: (301) 286-1617 \\
            {\tt silverberg@stars.gsfc.nasa.gov} }

\begin{abstract}
Observations from the first flight of the Medium Scale Anisotropy
Measurement (MSAM1-92) are analyzed to search for the Sunyaev-Zel'dovich (SZ)
effect towards the Coma cluster. 
This balloon-borne instrument uses a $28\arcmin$ FWHM beam 
and a three 
position chopping pattern with a throw of $\pm40\arcmin$. 
With spectral channels at 5.7, 9.3, 
16.5, and 22.6~\icm, the observations simultaneously sample
the frequency range where the SZ spectral distortion in the intensity 
transitions from a decrement to an increment and   
where the fractional
intensity change is substantially larger than 
in the Rayleigh-Jeans region. We set  limits on
the Comptonization parameter integrated over our antenna pattern, $\Delta 
y \leq 8.0 \times 10^{-5}$($2 \sigma$).  
For a spherically symmetric isothermal
model,  this implies a central
Comptonization parameter, $y_o \leq 2.0 \times 10^{-4}$, or a central 
electron density, $n_o \leq 5.8 \times10^{-3}$cm$^{-3}h_{50}$, a result
consistent with central densities implied by X-ray brightness measurements
and central Comptonization estimates from lower frequency observations 
of the SZ effect.
\end{abstract}

\setcounter{footnote}{0}

\section{Introduction}

Compton scattering of the Cosmic Microwave Background (CMB) photons by a 
hot electron gas results in 
a characteristic spectral signature,  the Sunyaev-Zel'dovich (SZ) effect 
(\cite{sunyaev72}). This effect is of 
cosmological interest 
because it serves as a 
probe of intracluster gas and its evolution and,
when combined with X-ray observations, can be used to
estimate the Hubble constant independent of the distance ladder
(\cite{birkinshaw92}, \cite{birkinshaw94b}). 
Nearby clusters, such as Coma, are particularly
interesting because high quality X-ray measurements exist (\cite{briel92},
\cite{white93}) allowing mapping of the gas distribution
in the cluster. 

Detection of the small change in the CMB intensity along lines
of sight through a hot gas-rich cluster has been a challenging task.  
The SZ effect has been detected in a number of clusters
(see \cite{birkinshaw94a}, \cite{rephaeli95a} for recent reviews of 
observations and theory) 
and even in the Coma cluster (\cite{herbig95}). 
At radio frequencies, the presence of
weak background sources may lead to significant systematic errors of 
either sign, canceling or mimicking the real SZ effect.  
Recently images  of the effect (\cite{jones93},
\cite{carlstrom96}) have been obtained using radio interferometric
arrays.
Early attempts to go to higher frequencies ($>$3 \icm) using 
ground-based instrumentation to take advantage of wide-band bolometric 
detectors have encountered 
problems with the atmosphere (\cite{meyer83}, \cite{radford86}, \cite{chase87}).
More recently, high quality ground-based measurements at millimeter wavelengths 
have been
obtained using difference techniques with an array of detectors 
(\cite{wilbanks94}).
In the interesting frequency range where the SZ distortion makes its
transition from a decrement to an increment, the atmosphere is not
very transparent; 
balloon-borne instruments allow observations at these frequencies 
and allow detection over wider
bandwidths
(\cite{page90}, and \cite{cheng94}, hereafter Paper I). 

In this letter we describe our attempt to 
detect the SZ effect in the Coma cluster during the first flight of the 
Medium Scale Anisotropy Measurement (MSAM1-92). We briefly describe the 
instrument, observations of the Coma cluster, and the data analysis 
procedures. Finally, we  compare these results with a model based on
parameters derived from X-ray observations of the Coma cluster.

\section{Instrument and Observations}

\cite{fixsen96} and Paper I provide a detailed description of the
instrument and observing method; here we briefly summarize only the 
essential features.
MSAM1 is a balloon-borne, 
off-axis Cassegrain telescope with
a 1.4~m primary mirror, a 0.27~m nutating secondary mirror, and a
$28\arcmin$ FWHM beam.  The 
four-frequency bolometric
radiometer operates at 0.24~K using a $^3$He refrigerator and
was previously flown in the Far Infra-Red Survey experiment 
(\cite{page90}, \cite{meyer91}).
Table~\ref{page_table} shows the effective frequencies and bandwidths 
of the four channels for a source with an SZ spectrum.
The bolometer output from each of the four channels is sampled 
synchronously with the secondary chopper  and is digitized at 32~Hz for 
real-time telemetry to the ground.
During the flight, MSAM1-92 scanned
Jupiter and Saturn to calibrate the instrument (calibration uncertainty
is estimated to be 10\%),
rastered over Jupiter to map the antenna beam,
scanned over the center of the Coma cluster X-ray source 
to search for the SZ effect, 
and integrated for 4.9~hours on a patch of sky near the North Celestial 
Pole to search for CMB anisotropy (Paper I).  

The Coma observations are done by pointing the telescope near the X-ray center
of the Coma cluster. The telescope is then moved 
$\pm 45\arcmin$ in cross-elevation with a period of 1 minute
causing the telescope beam to cross within $\sim1\arcmin$ of 
the X-ray 
center of the cluster at closest approach.
After scanning across the cluster for 2 minutes, a
nearby reference field $1\fdg05$ in elevation above the X-ray center was
scanned in the same manner for a total of 2 minutes.
After returning to reproduce the first set of scan across Coma for 
another 2 minutes, 
another off-source reference field $1\fdg05$ below the X-ray center was 
scanned for 2 minutes, followed by a 
final set of scans carried out for 4 minutes across the X-ray 
center. After each set of scans, motion was paused for 40~s to 
record an image with the onboard star camera to verify pointing. The total 
elapsed time for these observations was less than 20 minutes during 
which 8 minutes of data crossing the Coma cluster X-ray center were 
obtained.

Figure~\ref{star_cam} shows the scan pattern superimposed on the 
approximate star camera field of view 
at the time the Coma crossing observations were started. 
The circles show the
positions of (expected) stars that were detected by the star camera
at the start of the Coma central scans.
The relative location of the telescope beam in the 
camera frame and the instrument calibration are
fixed by the simultaneous observation of Jupiter with the star camera 
and the radiometer.  
Both the beam position and the calibration were confirmed near the end 
of the flight by a
similar observation of Saturn. 
Between images, the position is interpolated from
the gyroscope signals.
The resulting final pointing is accurate to 
$2\farcm5$, limited by  gyroscope drift.   

\section{Analysis}

\subsection{Level of Comptonization in Coma}

The level of Comptonization towards Coma may be estimated in the
context of a spherically symmetric, polytropic model for the intracluster 
gas. The electron density at a radial position, $r$, is taken to have 
the form (\cite{king66})
\beq
n_e(r)=n_{o}[1+(r/r_c)^2]^{-\frac{3}{2}\beta}
\label{isotherm}
\eeq
\noindent
where $r_c$ is the core radius of the cluster, $n_{o}$ is the central 
density, and $\beta$ is a density slope parameter. In polytropic models,
the gas temperature is proportional to $n^{\gamma -1}$, where $\gamma$ is the 
polytropic index.

The Comptonization parameter is 
\beq
y=(k\sigma_T/mc^2)\int n_e T_e dl,
\label{yy}
\eeq
where $k$ is the Boltzmann constant, $\sigma_T$ is the Thomson cross
section, $m$ is the electron mass, $c$ is the speed of light, 
$T_e = T_o [1+(r/r_c)^2]^{-(\frac{3}{2}\beta) (\gamma -1)}$ is the electron 
temperature, and the integral is along the line of sight. Substituting 
eq. \ref{isotherm} in eq. \ref{yy}, we obtain the Comptonization parameter at an angular 
distance $\theta$ from the center of the cluster
\beq
y(\theta)=Yf(\theta) , \hspace{0.3in} Y=\frac{2kT_o\sigma_Tn_e(0)r_c}{mc^2}
\eeq
\noindent
and
\beq
f(\theta)=(1+(\theta/\theta_c)^2)^{-(\xi-\frac{1}{2})} \int_0^{p} 
(1+t^2)^{-\xi} dt
\label{f_integ}
\eeq
where $\xi=\frac{3}{2}\beta \gamma$, $\theta_c$ is the angular core radius, 
$t$ is the distance from the center of the cluster in core radii, 
and $p$ is the outer radius of the gaseous
sphere in core radii.   The
\ROSAT\ results trace X-ray emission out to beyond $100\arcmin$. 
In contrast to the X-ray emission which depends quadratically on the
electron density, the SZ effect depends linearly on the electron density; 
we have, therefore, integrated to  $\infty$. Because the integral converges
rapidly, this has little effect on the result. The difference in intensities 
at a frequency, $\nu$, along different (pencil-beam) lines of sight is then 
(\cite{rephaeli95a})
\beq
\Delta I= \frac{2(kT)^3}{(hc)^2} g(x) \Delta y
\label{deltaI}
\eeq
\noindent
where
\beq
 g(x)=\frac{x^4e^x}{(e^x-1)^2} [x \coth(\frac{x}{2})-4],
\eeq
and 
\beq x=\frac{h\nu}{kT},
\eeq
$h$ is the Planck constant, and $T$ is the CMB temperature.   High precision
observations will have to take relativistic effects into account
at high frequencies even for $kT_o\sim5$KeV (\cite{rephaeli95b}).

\subsection{Results}

The MSAM1-92 configuration approximately samples the difference 
\[
\Delta y=
\left\{
\begin{array}{ll}
(y_l-y_r) & \mbox{(single-difference)}\\
y_c-\frac{1}{2} (y_l+y_r)  & \mbox{(double-difference)}
\end{array}
\right.
\]
where $y_c$, $y_l$, and $y_r$ are the expected Comptonization parameter in 
the central, left, and right beams, respectively.
Equation  \ref{deltaI} is integrated over the actual antenna pattern and over 
the channel bandpasses to estimate the observed 
intensity difference for a pure SZ spectrum source at each channel.

When the isothermal model described above is 
convolved with our beam maps, we find that our finite beam 
size and chopper 
throw give a response very similar to our point source response
(Jupiter scans).  The observed Comptonization parameter, $\Delta y$, is 
reduced from the central Comptonization of this
model by a factor
of $\sim$2.5(2.0) for the double(single)-difference data compared to a true
pencil beam measurement, $y_o$.  We correct for this geometric form 
factor later.
We simultaneously fit the Coma scan data to the point source response 
and a separate quadratic function of time
for each of the three central scans.  
This permits us to detect and remove any slow drifts from the 
time series in each channel and to determine the best fitting surface brightness
consistent with a source at the X-ray center of Coma.  The simultaneous
fit to the model and the baseline drift avoids removal of signal that may result
if separate fits were used sequentially.

The results of these fits to the data are summarized in 
Table~\ref{coma_limits}.
The procedure for estimating noise is not the same as we used in Paper I.
The majority of those observations were done near transit, while the Coma 
observations were done as Coma was setting, resulting in
a rapidly changing elevation during the observations.  Because of the short
time under these different observing conditions, we estimate
the noise on the Coma observations from
the dispersion in the data after removal of drifts.  Thus, the
reduced $\chi^2$ of all fits is unity.  
No significant detection is
found at any frequency, even when the 
position of the assumed X-ray center along the scan 
was allowed to vary from the known position.
Errors derived from the single-difference data are significantly
larger than for the double difference data. 
This may be due to the 
greater sensitivity of the single-difference to contamination from
atmospheric gradients.  The 22.5 \icm\ channel is expected to be
insensitive to an 
SZ spectral signature
and is primarily used as a diagnostic for possible dust contamination.
No significant contamination from warm dust is expected based on observations
of this region by \IRAS\ (\cite{wise93}), and none is observed. 
  
\section{Discussion}

To compare these limits with observations at other
wavelengths where different observing techniques and beam sizes are used, 
we use the model above to estimate the central Comptonization parameter 
and the corresponding central electron density limit.  
The isothermal gas parameters in Coma have been determined from 
an analysis of \ROSAT\ and \GINGA\ data (\cite{briel92}):
$kT_o=(8.2\pm 0.2)$ keV, $n_o=(2.89 \pm 0.04)\times 10^{-3}h_{50}^{1/2}$ cm$^{-3}$, 
$\beta=(0.75\pm0.03)$, and $\theta_c=(10.5\pm0.6)~\arcmin$. 
Using a redshift,
$z=0.023$ and $H_0$=50 km s$^{-1}$ Mpc$^{-1}$,
we estimate the central Comptonization 
parameter implied by the X-ray data is, $y_o=Y f(0)= (1.0\pm0.4)\times10^{-4}$ \label{eq_y0}. 
Modifying the observed results
by the form factors discussed above
gives $y_0 \leq 2.0 \times 10^{-4}$,  and 
$n_o \leq 5.8 \times10^{-3}$cm$^{-3}h_{50}$. These are consistent with 
the central density and degree of Comptonization estimated from the 
\ROSAT\ X-ray data and with the estimate for the central Comptonization of
$y_o=(9.3\pm1.7)\times10^{-5}$ made by \cite{herbig95} from measurements with
a  $7\arcmin$ beam and $22\arcmin$ beam separation
at 32 GHz.

Recent observations of the intrinsic CMB anisotropies on angular scales
$\sim$0\fdg5 (Paper I,
\cite{cheng96}, \cite{devlin94}, \cite{netterfield96}) are
showing $\Delta T/T$ $\sim
2\times 10^{-5}$.  
This intrinsic level
represents a potentially significant contaminant
to SZ observations at these angular scales because the CMB fluctuations 
may be comparable to the expected
signal from the SZ effect. 
 At higher frequencies (e.g. 16 \icm)
the shape of the SZ spectrum will aid in distinguishing SZ fluctuations
from intrinsic CMB fluctuations. 

Although the Coma cluster observations we present
represent only a relatively short observation during the 
first MSAM1-92 flight, the high sensitivity allows us to set limits on 
the Comptonization
which are comparable to the predicted value based on X-ray observations
and lower frequency measurements.
Increases in the sensitivity of bolometric detectors and better
observing strategies are
anticipated, leading to 
even higher quality measurements of the
SZ effect at these high ($\nu>5$\icm) frequencies.

We wish to thank the staff of the National Scientific Balloon
Facility in Palestine, Texas for their excellent support.  
We would like to thank J.~Mather, H.~Moseley, 
C.~Bennett, E. Dwek and M.~Hauser for helpful discussions.
The National Aeronautics and Space Administration supports this research 
through grants  NAGW 1841, RTOP 188-44, NGT 50908, and NGT 50720.
\singlespace

\begin{deluxetable}{ccccc}
\tablecolumns{5}
\tablecaption{Integrals$^{1}$ of Passbands over Sunyaev-Zel'dovich Spectrum
\label{page_table}}
\tablehead{
\colhead{Channel} 
&\colhead{1}&\colhead{2}&\colhead{3}&\colhead{4}\\
}
\startdata
Power ${\rm (Watts/(cm^{2}\,sr\,y))}$ 
	& $-3.36\times10^{-11}$
	& $8.36\times10^{-11}$
	& $7.0\times10^{-11}$
	& $1.1\times10^{-11}$\nl
$<\nu >$ [$\Delta\nu$] (\icm)
	& 5.52 [1.24]
	& 9.67 [2.12]
	& 16.2 [1.8]
	& 22.4 [1.2]\nl
Limit on out of band response $^{2}$
	& $<1$\%
	& $<1$\%
	& 4\%
	& 22\%\nl
\enddata
\tablecomments{
{1)}
The integrals for the SZ effect are 
normalized to y = 1. They are the passbands integrated over the
non-relativistic expression for the SZ effect in \cite{sunyaev72}.  
} 
\tablecomments{{2)} The leakage is $100\times $ ``out-of-band'' response
divided by ``in-band" response.
}
\end{deluxetable}

\begin{deluxetable}{ccccc}
\tablecolumns{5}
\tablecaption{Amplitudes of Coma Cluster Source  and Central Comptonization 
Parameter Estimates
\label{coma_limits}}
\tablehead{
&\colhead{$\Delta T_{RJ}$} 
&\colhead{$\Delta y$}&\colhead{Estimated}&\colhead{Central Comptonization}\\
\colhead{Channel}&&&\colhead{Form Factor$^1$}&\colhead{$y_o=kT_e/mc^2\int n_e dl$}\\
&\colhead{(\uK)}&\colhead{(x$10^{-4}$)}&&\colhead{(x$10^{-4}$)}
}
\startdata
\cutinhead{Single Difference}
1&-99$\pm$112&1.0$\pm$1.2&2.0&2.0$\pm$2.3\nl
2&17$\pm$49&0.34$\pm$1.0&2.0&0.68$\pm$2.0\nl
3&7$\pm$45&0.36$\pm$2.2&2.0&0.73$\pm$4.5\nl
4&-7$\pm$108&---&&---\nl
\tableline
Weighted Average&&0.59$\pm$0.71&&1.2$\pm$1.4\nl
\cutinhead{Double Difference}
1&50$\pm$43&-0.51$\pm$0.45&2.5&-1.3$\pm$1.1\nl
2&34$\pm$21&0.69$\pm$0.41&2.5&1.7$\pm$1.0\nl
3&23$\pm$20&1.2$\pm$1.0&2.5&2.9$\pm$2.5\nl
4&1$\pm$47&---&&---\nl
\tableline
Weighted Average&&0.22$\pm$0.29&&0.56$\pm$0.73\nl
\enddata
\tablecomments{ 1) The estimated form factor represents the factor by which 
the peak signal 
is reduced after the MSAM1 beam and chopping strategy are 
applied to a pure SZ source with 
parameters, $\theta_c=10.5\arcmin$, $\beta$=0.75 (\cite{briel92}).  
For this model, the MSAM1 chopper 
throw is not large enough for the offcenter beam(s) to be in a region of 
negligible contribution.}
\end{deluxetable}

\newpage
\figcaption{Coma observations scanning pattern. 
The crosses show the locations where field stars brighter than
$\sim$5th magnitude were located.  Circles are centered at the locations
where stars were actually detected by the star camera.  The contour map 
shows the X-ray data from the \ROSAT\ PSPC.   Lines shown are the scans
above, below and through the center of the Coma cluster.
The coordinates used are J1992.5. \label{star_cam} }

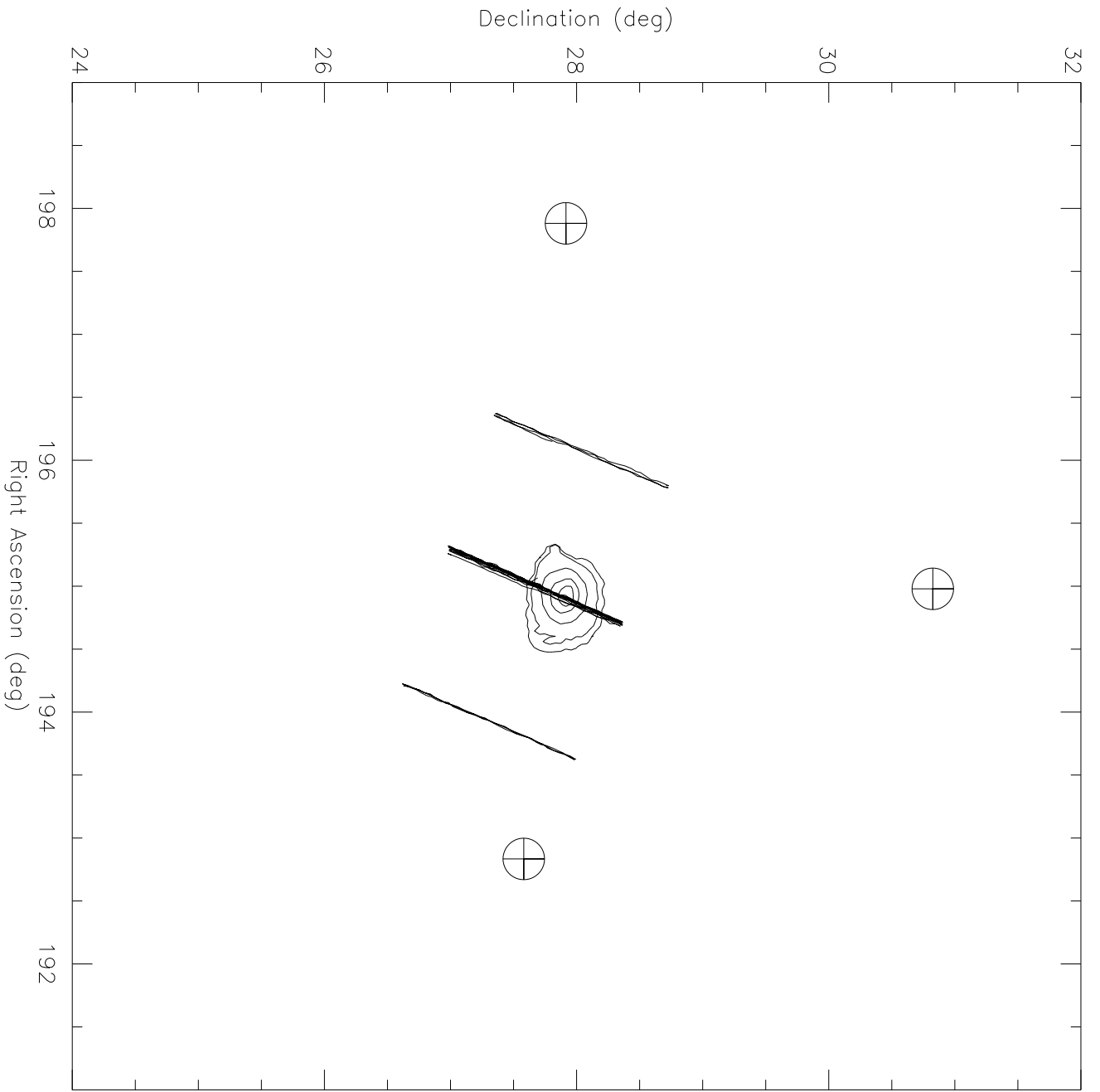
\end{document}